\documentclass[chaos, preprint, aip]{revtex4-1}
\usepackage{amsmath}
\usepackage{amsfonts}
\usepackage{amssymb}
\usepackage{dsfont}
\usepackage{float}
\usepackage{graphicx}
\usepackage{color}
\begin{document}

\title[]{Multiscale Dynamics in Communities of Phase Oscillators}

\author{Dustin Anderson}
\affiliation{
Department of Physics and Astronomy\\
Carleton College, Northfield, Minnesota 55057}
\author{Ari Tenzer}
\affiliation{
Department of Physics\\
Washington University in St. Louis, St. Louis, MO 63105}
\author{Gilad Barlev}
\author{Michelle Girvan}
\author{Thomas M. Antonsen}
\author{Edward Ott}
\affiliation{ 
Institute for Research in Electronics and Applied Physics \\
University of Maryland, College Park, Maryland 20742
}%

\date{\today}
             
\begin{abstract}
We investigate the dynamics of systems of many coupled phase oscillators with heterogeneous frequencies. We suppose that the oscillators occur in $M$ groups.  Each oscillator is connected to other oscillators in its group with ``attractive'' coupling, such that the coupling promotes synchronization within the group.  The coupling between oscillators in different groups is ``repulsive"; i.e., their oscillation phases repel.  To address this problem, we reduce the governing equations to a lower-dimensional form via the ansatz of Ott and Antonsen \cite{ref1}.  We first consider the symmetric case where all group parameters are the same, and the attractive and repulsive coupling are also the same for each of the $M$ groups.  We find a manifold $\mathcal{L}$ of neutrally stable equilibria, and we show that all other equilibria are unstable.  For $M \ge 3$, $\mathcal{L}$ has dimension $M-2$, and for $M = 2$ it has dimension $1$.  To address the general asymmetric case, we then introduce small deviations from symmetry in the group and coupling parameters.  Doing a slow/fast timescale analysis, we obtain slow time evolution equations for the motion of the $M$ groups on the manifold $\mathcal{L}$.  We use these equations to study the dynamics of the groups and compare the results with numerical simulations.

\end{abstract}

\pacs{}
\keywords{}
\maketitle
\textbf{The dynamics of hierarchically structured networks is particularly relevant to systems with very large numbers of dynamical units.  Examples can be found in neuroscience and in the study of social dynamics in large populations.  At the coarsest level, one can often think of such systems as being based on interactions between distinct groups or communities of dynamical network nodes.  Thus, it is of great interest to investigate the kinds of dynamical behaviors that can result when network systems have interacting communities, and to develop techniques that may be useful for studying such systems.  In this paper we address these issues for the illustrative, paradigmatic case in which the nodal dynamics is describable within the phase oscillator model.}

\section{Introduction}
\label{sec:introduction}
It is often observed that real-world networks have groups of nodes that are associated strongly with each other and less strongly with nodes in other groups\cite{ref2, ref3}.  Considering situations in which the links on such networks represent the flow of signals between interacting nodal dynamical units, this type of group structure may fundamentally influence the functioning, overall performance, and emergent dynamical properties of such systems.  In this paper we investigate the dynamics of a simple model for the interaction of groups of many oscillators.  Following Kuramoto, there is an oscillator at each network node whose state is given purely by its oscillation phase angle\cite{ref4, ref5, ref6, ref7, ref8}.  The oscillators occur in $M$ groups and have different natural oscillation frequencies.  Every oscillator is uniformly and ``attractively'' coupled to all other oscillators in its group, but is also ``repulsively'' coupled to all oscillators outside its group.  By an ``attractive'' (``repulsive'') network coupling, we mean that the coupling promotes (works against) phase synchronization of the two oscillators that are linked by the coupling. The general problem of interacting network oscillator groups has also been considered\cite{ref1, ref9}, and the nonlinear dynamics of such groups has been studied for the case of groups whose average natural frequencies are not in resonance \cite{ref10} .  Our focus on within-group attraction and between-group repulsion is partly motivated by our expectation that this will provide a particularly clear illustration of the effect of network group structure on system dynamics.  If we make a loose analogy between the phase oscillator model and opinion dynamics, as has been done in the past\cite{ref11}, we can think of the interactions in the context of a multi-party political system, with within-group attractions representing the tendencies for individuals to align with those in the same political party and between-group repulsions reflecting the desire of individuals to differentiate themselves from members of other parties.

We consider our model using direct numerical simulations and by the low-dimensional reduction (Sec.\ \ref{sec:lowDimensional}) of Ott and Antonsen \cite{ref1, ref12}, which has proven useful for the treatment of a wide variety of problems involving the interactions of a large number of phase oscillators \cite{ref13}.  We first consider a symmetric situation (Sec.\ \ref{sec:identicalGroups}), in which all parameters describing the $M$ oscillator groups are identical, and the attractive and repulsive couplings within and between the groups is uniform.  It is found, both numerically and analytically, that, above a threshold in the coupling, the oscillations within groups may become coherent\cite{ref14}, and the groups interact to achieve equilibrium configurations.  We also analytically determine all possible system equilibria and find that, when $M \ge 3$,  there is an ($M-2$) dimensional (one dimensional for $M = 2$) manifold $\mathcal{L}$ of possible neutrally stable equilibria, and that all other equilibria are unstable.  In order to provide insight into the general asymmetric case (Sec.\ \ref{sec: nonidenticalGroups}), we introduce small asymmetric deviations to the group and coupling parameters.  We then analyze the situation using multiple-timescale asymptotics\cite{ref15}. This results in a set of equations describing the slow-timescale evolution of the $M$ group order parameters as they temporally evolve and interact on the manifold $\mathcal{L}$ of neutrally stable symmetric solutions.  We also report numerical results testing the applicability of our results.  Further discussion and conclusion are given in Sec.\ \ref{sec:conclusion}.

\section{Low Dimensional Formulation}
\label{sec:lowDimensional}

In Kuramoto's original formulation, the dynamics of the individual oscillators are given by 
\begin{equation} \label{eq: kuramoto}
\frac{d\theta_i(t)}{dt} = \omega_i + \frac{K}{N}\sum_{j=1}^N \sin[\theta_j (t) - \theta_i (t)],
\end{equation}
where $\theta_i$ is the phase of the $i$th oscillator ($i = 1, 2, ..., N$), $\omega_i$ is the $i$th oscillator's natural frequency, and $K$ measures the strength of coupling between oscillators.  We consider a modified version of this system in which the oscillators are placed into $M$ communities (groups) of equal size
.  Instead of a single coupling constant, we consider a matrix in which the $\sigma \sigma'$th element represents the coupling of each oscillator in group $\sigma$ ($\sigma = 1, 2, ..., M$) to each oscillator in group $\sigma'$. In this model, the dynamics of the oscillators are governed by 
\begin{equation} \label{eq: groups}
\frac{d\theta_i^{\sigma}}{dt} = \omega_i^{\sigma} + \sum_{\sigma'=1}^M \frac{M K_{\sigma \sigma'}}{N} \sum_{j = 1}^{N/M} \sin(\theta_j^{\sigma'} - \theta_i^{\sigma}),
\end{equation}
where $\theta_i^{\sigma}$ is the phase of the $i$th oscillator in group $\sigma$.  The natural frequencies $\omega_i^{\sigma}$ are taken from Lorentzian distributions given by
\begin{equation}
\label{eq: lorentzians}
g^\sigma(\omega) = \frac{\Delta_\sigma /\pi}{(\omega - \bar{\omega}_\sigma)^2 + \Delta_\sigma^2},
\end{equation}
where $\Delta_\sigma$ and $\bar{\omega}_\sigma$ are the width and center, respectively, of the Lorentzian distribution for group $\sigma$.

Just as the dynamics of the original Kuramoto model can be captured by a single complex order parameter, the modified system can be understood through the behavior of the $M$ group order parameters
\begin{equation} \label{eq: orderParams}
\alpha_\sigma = \frac{M}{N}\sum_{i = 1}^{N/M} \exp(i \theta_i^{\sigma}).
\end{equation}
As shown by Ott and Antonsen\cite{ref12}, the group order parameters are attracted to a manifold on which they evolve according to the ordinary differential equations (ODEs)
\begin{equation}
\label{eq: lowdimeq}
\frac{d\alpha_\sigma}{dt} = -(\Delta_\sigma - i \bar{\omega}_\sigma)\alpha_\sigma - \frac{1}{2} \sum_{\sigma' = 1}^M K_{\sigma \sigma'}[\alpha_{\sigma'}^* \alpha_{\sigma}^2 - \alpha_{\sigma'}],
\end{equation}
where the $^*$ denotes complex conjugation.
Defining 
\begin{equation}
\label{eq: R}
R_\sigma = \sum_{\sigma' = 1}^M K_{\sigma \sigma'} \alpha_{\sigma'},
\end{equation}
we rewrite Eq.\ \eqref{eq: lowdimeq} as
\begin{equation}
\label{eq: lowdim2}
\frac{d\alpha_\sigma}{dt} = -(\Delta_\sigma - i \bar{\omega}_\sigma)\alpha_\sigma - \frac{1}{2} (\alpha_{\sigma}^2 R_\sigma^* - R_\sigma).
\end{equation}

\section{Identical Groups}
\label{sec:identicalGroups}

\subsection{Equilibria}
\label{subsec: equilibria}

We first consider the simplest case, in which $\Delta_\sigma = 1$ and $\bar{\omega}_\sigma = \bar{\omega}$ for all $\sigma$.  In order to investigate the effects of repulsion between groups of oscillators, we define the entries of the group coupling matrix to be
\begin{equation}
\label{eq: Ks}
K_{\sigma \sigma'} = \left\{
     \begin{array}{lr}
       K & : \sigma = \sigma'\\
       -K b & : \sigma \not = \sigma'\\
     \end{array}
   \right.
\end{equation}
where the constant $K > 0$ represents the overall coupling strength, and $b > 0$ quantifies the degree of repulsion between oscillators in different groups.  The average frequency $\bar{\omega}$ can be transformed to zero by the change of variables $\theta_i \to \theta_i - \bar{\omega} t$, and we henceforth set $\bar{\omega} = 0$.  [In Section \ref{sec: nonidenticalGroups} we will consider deviations from $\Delta_\sigma = 1$, $\omega_\sigma = 0$ and Eq.\ \eqref{eq: Ks}.]  Defining 
\begin{equation}
\label{eq: S}
S = \sum_{\sigma = 1}^M \alpha_\sigma,
\end{equation}
Eq.\ \eqref{eq: R} can be expressed as
\begin{equation}
\label{eq: R2}
R_\sigma = K(1+b)\alpha_\sigma - K b S,
\end{equation}
and Eq.\ \eqref{eq: lowdim2} becomes
\begin{equation}
\label{eq: lowdim3}
\frac{d\alpha_\sigma}{dt} = -\alpha_\sigma - \frac{K}{2} \left[\alpha_{\sigma}^2 \left((1+b)\alpha_\sigma^* - b S^*\right) - \left((1+b)\alpha_\sigma - b S\right)\right].
\end{equation}
In order to characterize possible equilibria of the system, we set $d\alpha_\sigma/dt = 0$.  Writing $\alpha_\sigma$ and $S$ in polar form as $r_\sigma e^{i \psi_\sigma}$ and $S_0 e^{i \Psi}$, respectively, we substitute these quantities into Eq.\ \eqref{eq: lowdim3} and rearrange terms to yield
\begin{equation}
\label{eq: lowdim4}
0 = -r_\sigma - \frac{K}{2}\left[r_\sigma(1+b)(r_\sigma^2 - 1) - r_\sigma^2 b S_0 e^{i(\psi_\sigma - \Psi)} + b S_0 e^{-i(\psi_\sigma - \Psi)}\right].
\end{equation}
Taking the imaginary part of both sides yields
\begin{equation}
\label{eq: lowdim5}
0 = \frac{K}{2} b S_0 (r_\sigma^2 + 1) \sin(\psi_\sigma - \Psi).
\end{equation}
This equation gives restrictions on the possible steady-state equilibria in our model.  Assuming $K \not = 0$ and $b \not = 0$, Eq.\ \eqref{eq: lowdim5} is satisfied if and only if either $S_0 = 0$, or $\psi_\sigma - \Psi = k \pi$ for some integer $k$.  The latter condition must be satisfied for $\sigma = 1, 2, ..., M$, implying that, for an equilibrium with $S \not = 0$, the complex order parameters $\alpha_\sigma$ all lie on a single line through the origin of the complex $\alpha$-plane.

\subsection{Equilibria with $S = 0$}
\label{subsec: szero}
We now turn our attention to the equilibria with $S = 0$.  In this case, Eq.\ \eqref{eq: lowdim4} becomes
\begin{equation}
\label{eq: lowdim6}
0 = r_\sigma \left[1 + \frac{K}{2}(1+b)(r_\sigma^2 - 1)\right].
\end{equation}
Solving for $r_\sigma$ yields either $r_\sigma = 0$ or 
\begin{equation}
\label{eq: rsquared}
r_\sigma = r_0 \equiv \left[1 - \frac{2}{K(1+b)}\right]^{1/2}.
\end{equation}

In Kuramoto's original model, there existed a critical coupling value $K_c$ that marked the onset of synchronization in the system.  For our case, Eq.\ \eqref{eq: rsquared} yields an analogous value,
\begin{equation}
\label{eq: kc}
K_c = \frac{2}{1+b}.
\end{equation}
For $K < K_c$, the oscillators behave incoherently [$r_\sigma = 0$ is the only solution of Eq.\ \eqref{eq: lowdim6}] and exhibit no collective behavior.  From now on we assume that $K > K_c$, so that the oscillators in each group can synchronize with one another, and the groups act as coherent entities.  

\begin{figure}[h]
\centering
\includegraphics[width=12cm]{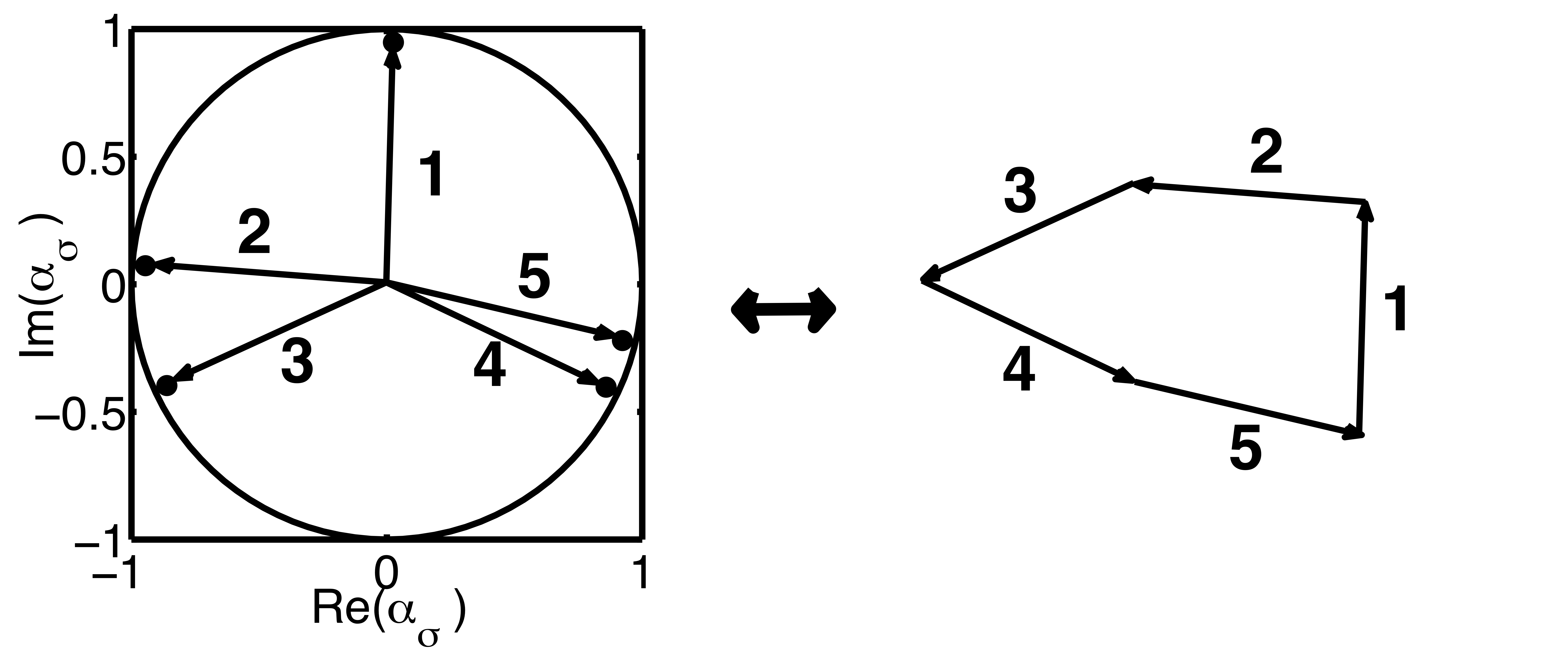}
\caption{\footnotesize{An $S=0$ equilibrium with five groups, and the corresponding equilateral pentagon. $|\alpha_\sigma| = r_0 < 1; \sigma = 1, 2, 3, 4, 5.$}}
\label{fig: Figure1}
\end{figure}

For $K > K_c$, an equilibrium with $S = 0$ can have some group order parameters with length zero and some with length $r_0$. It is convenient to visualize the nonzero order parameters as lying on a circle of radius $r_0$ in the complex $\alpha$-plane. Assume that there are $m$ groups with $r_\sigma = 0$ and $M-m$ groups with $r_\sigma = r_0$.  If we conceptualize the non-zero order parameters as vectors of length $r_0$, we can place them tip-to-tail to form an equilateral polygon in the plane.  Therefore, given $M-m$ groups with $r_\sigma = r_0$, the set of possible $S = 0$ equilibria corresponds to the set of equilateral $(M-m)$-gons.  For $(M-m) = 2$, the two $r_\sigma = r_0$ groups must have group phases that are diametrically opposed (i.e., separated by $\pi$).  If $(M-m) = 3$, the phases of the order parameters must be equally spaced by $2 \pi / 3$, corresponding to an equilateral triangle in the tip-to-tail representation.  If $(M-m) = 4$, the tip-to-tail configuration of order parameter vectors corresponds to the sides of a rhombus.  Since a rhombus has two pairs of parallel sides, there must be two pairs of diametrically opposed order parameters, but the angle between them is arbitrary (corresponding to the one parameter family of all possible rhombus shapes).  For $(M-m) = 5$ and larger, there is more freedom in the specification of equilateral $(M-m)$-gons (see Fig.\ \ref{fig: Figure1}), and the pattern is not as easily specified as for $(M-m) = 2$, $3$ and $4$.  In general, assuming $(M-m) \ge 3$, the set of all possible $(M-m)$-gon shapes, (and hence the family of all possible equilibria) is $(M-m-2)$ dimensional for $(M-m) \ge 3$.  That is, these equilibria lie on an $(M-m-2)$-dimensional manifold.  We denote this manifold by $\mathcal{L}$.  The case $(M-m) = 2$ is special; in this case, $\mathcal{L}$ has dimension one.  For $(M-m) = 2$ or $3$, the dimension of $\mathcal{L}$ is one, corresponding to invariance of these equilibria under a rigid rotation of the phases of $\alpha_\sigma$.

\subsection{Equilibria with $S \not = 0$}
\label{subsec: snonzero}
According to Eq.\ \eqref{eq: lowdim5}, any solution that does not have $S = 0$ must have all of the order parameters on a single line through the origin.  Without loss of generality, we can rotate the configuration so that the order parameters all lie on the real line, and the value of $S$ is positive.  Thus we can let $\psi_\sigma = \Psi = 0$ for all $\sigma$, if we allow $r_\sigma$ to take on both positive and negative real values.  Then Eq.\ \eqref{eq: lowdim4} becomes
\begin{equation}
\label{eq: snonzero}
0 = \frac{K}{2} (1+b) r_\sigma^3 - \frac{K}{2} b S r_\sigma^2 + \left(1 - \frac{K}{2} (1+b)\right) r_\sigma + \frac{K}{2} b S.
\end{equation}
The solutions to this cubic equation give three possible values for $r_\sigma$.  We note that since $S > 0$, the product of the roots of this equation, equal to $-b S/(1+b)$, is negative.  If the cubic has two complex roots, the third root must be negative to satisfy this condition.  Then, since the order parameters lie on the real line by assumption, all the order parameters must be at the negative real root.  But $S$ cannot be positive if all the order parameters have a negative value, so we reach a contradiction.  Therefore, we conclude that all of the roots of this equation must be real.  Furthermore, exactly one of the roots must be negative.

\begin{figure}[h]
\centering
\includegraphics[width=8cm]{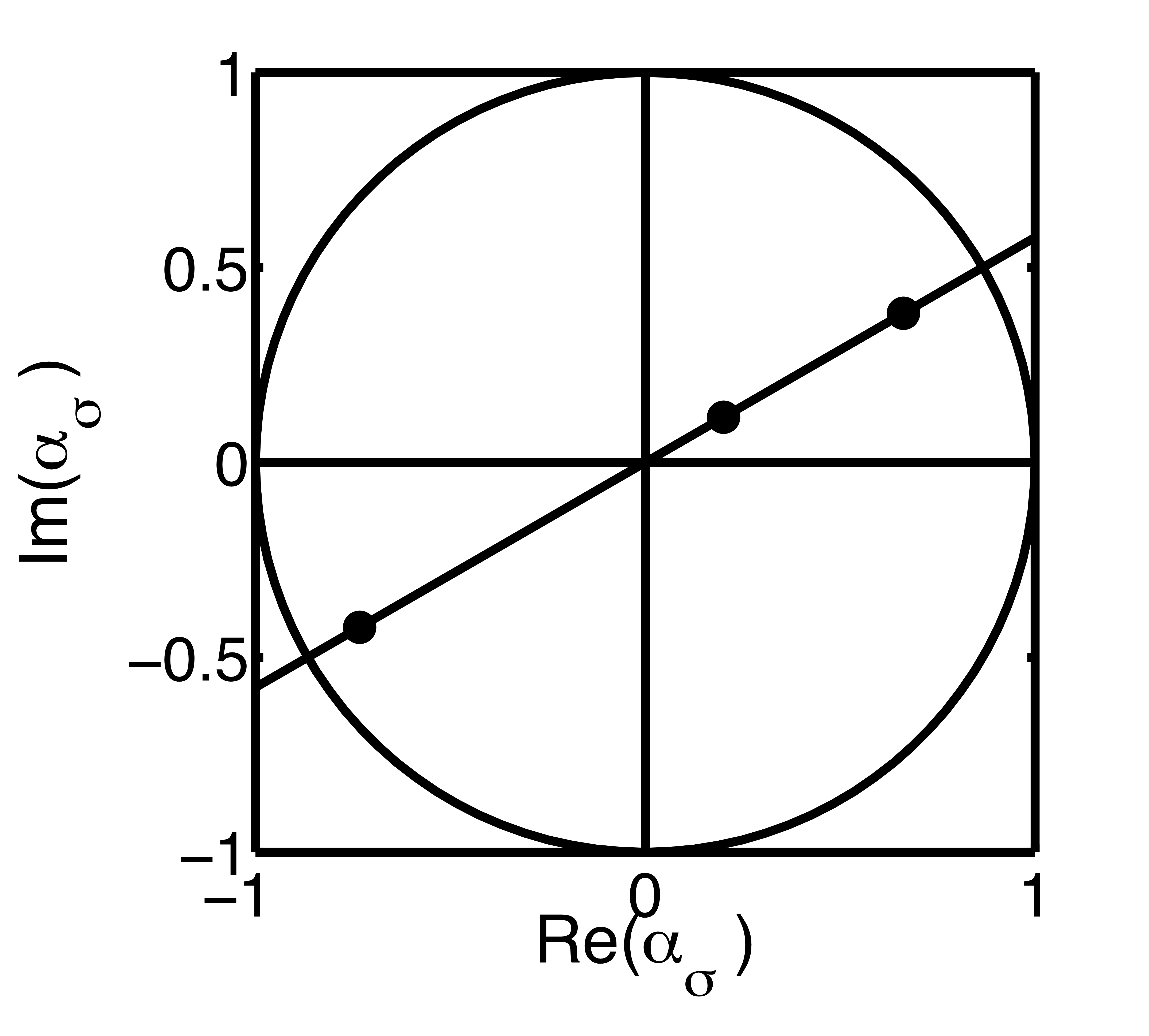}
\caption{\footnotesize{An equilibrium with $S \neq 0$.  The group order parameters lie on a straight line through the origin, and their magnitudes are all solutions to Eq.\ \eqref{eq: snonzero}.}}
\label{fig: Figure2}
\end{figure}

\subsection{Numerical Simulations}
\label{subsec: simulations}
In order to test the validity of these results, we integrated Eq.\ \eqref{eq: lowdim3} numerically for each oscillator group.  We chose values of $K$ and $b$ such that $K > K_c$, and examined the behavior of the order parameters for different numbers of groups.  We observed that, given arbitrary initial conditions for the group order parameters, the system tended to a approach a steady-state equilibrium.  All typical initial conditions yielded evolutions that tended toward $S = 0$ equilibria in which the lengths of all the order parameters were equal. If we chose the initial order parameters so that they lay on a line through the origin and satisfied Eq.\ \eqref{eq: snonzero}, we obtained stationary states with $S \not = 0$.  If some of the initial order parameters were zero and the initial configuration satisfied $S=0$, then we also obtained stationary states.  However, these configurations were seen to be unstable; any small changes in the order parameters caused the system to drift away and evolve towards a state with $S=0$ and all $r_\sigma = r_0$.  This implies that the only stable states of the system are those in which $S = 0$ and all $r_\sigma = r_0$.  In Sec.\ \ref{subsec: stability}, we demonstrate this analytically.

\begin{figure}[h]
\centering
\includegraphics[width=8cm]{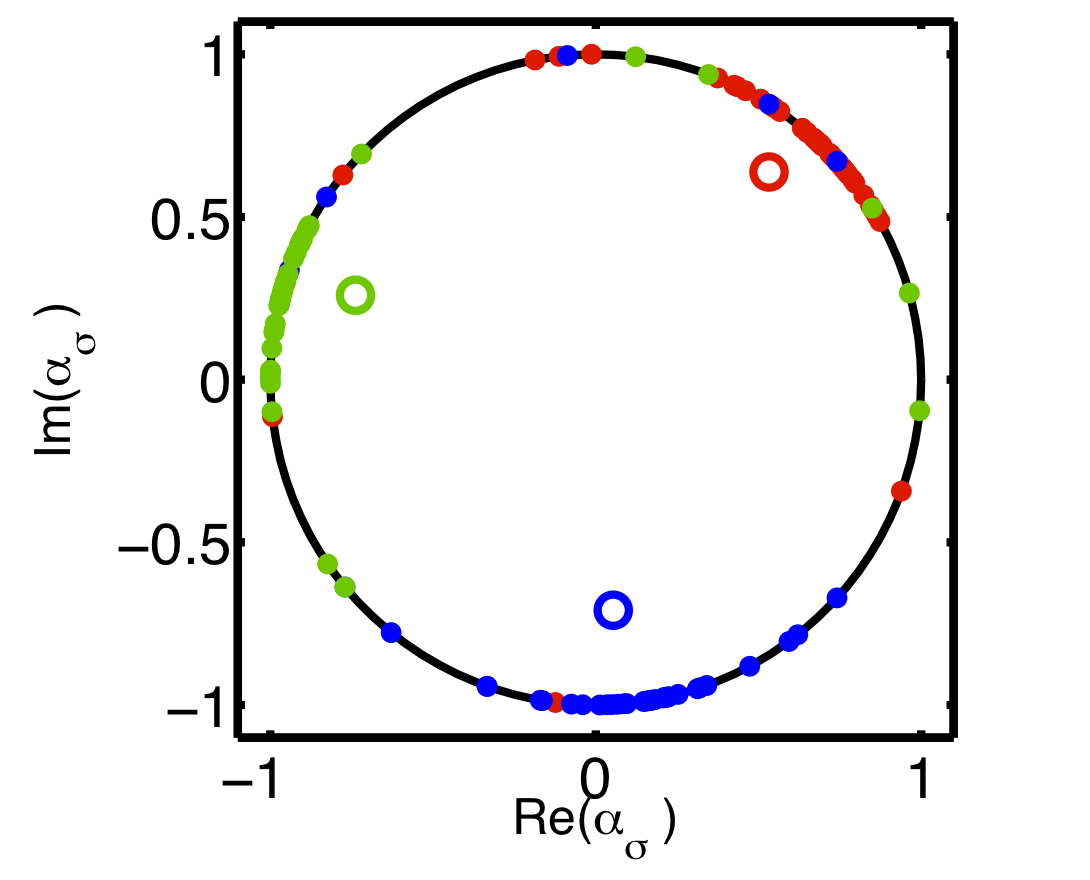}
\caption{\footnotesize{A sampling of oscillator phases from a numerical simulation with $M=3$.  Individual oscillator phases $\theta_i^\sigma$ appear as solid dots on the unit circle, and the colors (red, green and blue) correspond to different oscillator groups.  Group order parameters $\alpha_\sigma$ appear as larger open circles.}}
\label{fig: Figure3}
\end{figure}

We also conducted simulations of the full collection of oscillators by integrating Eq.\ \eqref{eq: groups} for each oscillator, using a group size of 2000 oscillators.  When $K > K_c$, we observed that the oscillators resolved into coherent groups with non-zero order parameters of nearly equal magnitude.  The group phases tended to arrange themselves such that $S \approx 0$.  However, because the number of oscillators was finite and their natural frequencies were randomly distributed, the average natural frequencies of the groups were not all exactly the same.  As a result of this asymmetry, the system did not reach a steady-state equilibrium.  Instead, the group order parameters rotated slowly in the complex plane at varying speeds.  The configuration changed continuously, but at every instant of time the system resembled an $S=0$ equilibrium with all $r_\sigma = r_0$.  As discussed above, the $S=0$ condition restricts the possible configurations of the system.  For $M=2$ or $3$, the phases of all group order parameters changed at the same speed, in order to keep the phases evenly spaced on the unit circle. For $M = 4$, the groups formed two pairs whose phases changed at different speeds.  For $M \ge 5$, the group phases evolved in a more complicated manner, reflecting the multitude of possible $S=0$ configurations for large numbers of groups.  An example with $M = 3$ is given in Fig.\ \ref{fig: Figure3}.

The absence of stationary states in these simulations is due to the asymmetries that naturally arise in finite collections of oscillators.  Our low-dimensional analysis of the identical-groups case does not account for these asymmetries.  In Section \ref{sec: nonidenticalGroups} we will explain this behavior by examining the low-dimensional system in the case in which the groups do not have identical properties.  

\subsection{Stability of Equilibria}
\label{subsec: stability}
\subsubsection{Equilibria with $S = 0$ and $r_\sigma \neq 0$ for all $\sigma$}
\label{subsubsec: szero}
To analyze the stability of the $S=0$ equilibria with $r_\sigma \not = 0$ for all $\sigma$, we introduce a small complex perturbation to each order parameter and determine whether the perturbations grow or shrink with time.  We replace $\alpha_\sigma$ with $r_0 e^{i \psi_\sigma}(1 + \delta \alpha_\sigma)$ in Eq.\ \eqref{eq: lowdim3} and replace $S$ with $S_0 + \delta S$, where $\delta S = \sum_{\sigma = 1}^M r_0 e^{i \psi_\sigma} \delta \alpha_\sigma$.  Letting $S_0 = 0$ and collecting terms that are first order in $\delta \alpha_\sigma$, we obtain
\begin{equation}
\label{eq: lowdimpert}
\frac{d(\delta \alpha_\sigma)}{dt} = -\delta \alpha_\sigma - \frac{K}{2}\left[(1+b)(r_0^2 \delta \alpha_\sigma^* + 2 r_0^2 \delta \alpha_\sigma) - b r_0 e^{i \psi_\sigma} \delta S^* - (1+b) \delta \alpha_\sigma + \frac{b}{r_0} e^{-i \psi_\sigma} \delta S\right].
\end{equation}
If we define the vector quantities,
\begin{equation}
\label{eq: vectors}
\vec{\delta \alpha} = \left( \begin{array}{c}
\delta \alpha_1 \\
\delta \alpha_2 \\
\vdots  \\
\delta \alpha_M \end{array} \right) \mbox{   and   }  
\vec \xi = \left( \begin{array}{c}
e^{i \psi_1} \\
e^{i \psi_2} \\
\vdots  \\
e^{i \psi_M} \end{array} \right),
\end{equation}
then we can cast the $M$ ODEs in Eq.\ \eqref{eq: lowdimpert} into the single vector equation,
\begin{equation}
\label{eq: lowdimmat}
\frac{2}{K} \frac{d(\vec{\delta \alpha})}{dt} = -\left[\left(r_0^2 (1+b)\right) \mathds{1} + b \left(\vec \xi^* \vec \xi^T\right)\right]\vec{\delta \alpha} - r_0^2\left[(1+b) \mathds{1} - b \left(\vec \xi \vec \xi^{T*}\right)\right]\vec{\delta \alpha}^*,
\end{equation}
where $\mathds{1}$ is the $M\times M$ identity matrix, $T$ denotes transposition, and we have made use of Eq.\ \eqref{eq: rsquared} to simplify the matrix multiplying $\vec{\delta \alpha}$.  The conjugate of Eq.\ \eqref{eq: lowdimmat}  is valid as well; those two equations can be combined into a single vector equation,
\begin{equation}
\label{eq: lowdimdouble}
\frac{d \vec{a}}{dt} = A \vec{a},
\end{equation}
where
\begin{equation}
\label{eq: AandB}
\vec{a} =
\left( \begin{array}{c}
\vec{\delta \alpha} \\
\vec{\delta \alpha}^* \end{array} \right),
A = -\frac{K}{2} \left( \begin{array}{cc}
r_0^2 (1+b) \mathds{1} + b \left(\vec \xi^* \vec \xi^T\right) &
r_0^2\left[(1+b) \mathds{1} - b \left(\vec \xi \vec \xi^{T*}\right)\right] \\
r_0^2\left[(1+b) \mathds{1} - b \left(\vec \xi^* \vec \xi^T\right)\right] &
r_0^2 (1+b) \mathds{1} + b \left(\vec \xi \vec \xi^{T*}\right)
\end{array} \right).
\end{equation}
We assume an exponential solution $\vec{a} \sim \exp{\lambda t}$, so that $d \vec{a}/dt = \lambda \vec{a}$ for some scalar $\lambda$.  Equation \eqref{eq: lowdimdouble} is then satisfied if $\lambda$ is an eigenvalue of the matrix $A$.  We can analyze the behavior of the system near equilibrium by finding the $2M$ eigenvalues of $A$ and determining whether their real parts are positive, negative or zero.

We begin searching for these eigenvalues by finding real vectors $\vec u$ satisfying $\vec \xi^T \vec u = 0$.  For such a vector $\vec u$, we see that both
\begin{equation}
\label{eq: uu}
       \vec{a}_+ = 
\left( \begin{array}{c}
\vec u \\
\vec u
\end{array} \right)
\mbox{   and   }       
        \vec{a}_- = 
\left( \begin{array}{c}
i \vec u \\
-i \vec u
\end{array} \right)
\end{equation}
are eigenvectors of matrix $A$ with eigenvalues $-[K(1+b)-2]$ and $0$, respectively.  The first of these eigenvalues is negative provided that $K > K_c$, which we have assumed.  The eigenvalue $0$ corresponds to neutrally stable perturbations; if $\vec{a}$ is parallel to an eigenvector with eigenvalue $0$, then the perturbation will neither grow nor decay in time.

Noting that $\delta S$ can be rewritten as $r_0 \vec \xi^T \vec{\delta \alpha}$, we see that the above condition on $\vec u$ is equivalent to requiring $\delta S = 0$.  Further, since $\vec u$ has real components, the order parameters $r_0 e^{i \psi_\sigma}(1 + \delta \alpha_\sigma)$ are such that only their magnitudes are perturbed in the case of $\vec{a}_+$, and only their phases are perturbed in the case of $\vec{a}_-$, where $\vec{a}_+$ and $\vec{a}_-$ are as given in \eqref{eq: uu}.  Because the eigenvalue corresponding to $\vec{a}_+$ is negative, perturbations leaving $S=0$ and changing only the magnitudes of the order parameters will decay exponentially in time.  The eigenvalue corresponding to $\vec{a}_-$ is zero, so perturbations leaving $S=0$ and changing only the angles of the order parameters will be neutrally stable.

Because $\vec \xi = \vec c + i \vec s$, where
\begin{equation}
\label{eq: cands}
\vec c = \left( \begin{array}{c}
\cos(\psi_1) \\
\cos(\psi_2) \\
\vdots  \\
\cos(\psi_M) \end{array} \right) \mbox{   and   }  
\vec s = \left( \begin{array}{c}
\sin(\psi_1) \\
\sin(\psi_2) \\
\vdots  \\
\sin(\psi_M) \end{array} \right),
\end{equation}
the real vectors $\vec u$ satisfying $\vec \xi^T \vec u = 0$ are exactly the vectors orthogonal to both $\vec c$ and $\vec s$.  For $M\ge 3$ this condition represents two constraints on $\vec u$, so the space of vectors $\vec u$ is $M-2$ dimensional.  Therefore there are $M-2$ independent eigenvectors of matrix $A$ with eigenvalue $0$, and $M-2$ with eigenvalue $-(K(1+b)-2)$.  For $M=2$, we have that $\psi_2 = \psi_1 + \pi$.  In this case the real and imaginary parts of $\vec \xi^T \vec u = 0$ give the same constraint, so the space of vectors $\vec u$ is one dimensional.

Since $A$ is a $2M\times 2M$ matrix, our analysis accounts for all but four of the $2M$ eigenvalues of $A$.  The previously identified eigenvectors span the space orthogonal to 
\begin{equation}
\label{eq: cossinvec}
\left( \begin{array}{c}
\vec c \\
\vec s
\end{array} \right),
\left( \begin{array}{c}
\vec c \\
-\vec s
\end{array} \right),
\left( \begin{array}{c}
\vec s \\
\vec c
\end{array} \right) \mbox{   and   }
\left( \begin{array}{c}
\vec s \\
-\vec c
\end{array} \right).
\end{equation}
The matrix $A$ must leave the space spanned by these four vectors invariant, because, for any $M$ dimensional vector $\vec v$, $\vec \xi \vec \xi^{*T} \vec v$ and $\vec \xi^* \vec \xi^{T} \vec v$ lie within that space.  We can thus consider the restriction of $A$ to that space, obtained by the transformation $V^{-1} A V$, where the $2M \times 4$ matrix $V$ is
\begin{equation}
\label{eq: v}
V = \left(\begin{array}{cccc}
\vec c & \vec c & \vec s & \vec s \\
\vec s & -\vec s & \vec c & -\vec c
\end{array} \right)
\end{equation}
and $V^{-1}$ is a left inverse of $V$ such that $V^{-1}V$ is the $4 \times 4$ identity matrix.  $V^{-1} A V$ is a $4\times 4$ matrix whose eigenvalues are the eigenvalues of $A$ that have corresponding eigenvectors in the restricted space.  Using Mathematica, we find that these four eigenvalues are the solutions to the two quadratic equations,
\begin{equation}
\label{eq: 4eigvals}
\lambda^2 +B_\pm \lambda + C_\pm = 0,
\end{equation}
where
\begin{equation}
\label{eq: Pdefn}
B_\pm = (K(1+b)-2) + \frac{K b}{2} (M \pm P r_0^2),
\end{equation}
\begin{equation}
\label{eq: Pdefn}
C_\pm = \frac{K b}{2} (1+r_0^2)(M \pm P)\left(\frac{K}{2} (1+b) - 1\right),
\end{equation}
\begin{equation}
\label{eq: Pdefn}
P = \sqrt{(\vec{c}^T \vec{c} - \vec{s}^T \vec{s})^2 + 4(\vec{c}^T \vec{s})^2} = \sqrt{\sum_{i = 1}^M \sum_{j = 1}^M \cos[2(\psi_j - \psi_i)]}.
\end{equation}
Note that $0 \le P \le M$.  For all four roots of these quadratics to be real and negative, we need both $B_\pm$ and $C_\pm$ to be positive, and $B_\pm^2 \ge 4C_\pm$.  Since $B_\pm$ is the sum of two positive numbers whenever $K \ge K_c$, it must be positive.  Additionally, $C_\pm$ is the product of five positive terms whenever $K \ge K_c$, so it must be positive.  In Appendix \ref{app: demon} we show that $B_\pm^2 \ge 4C_\pm$ as well.

We have thus established that the four remaining eigenvalues of matrix $A$ are negative for all valid choices of the system parameters.  Note that these eigenvalues correspond to perturbations $\vec{\delta \alpha}$ that are linear combinations of the vectors $\vec s$ and $\vec c$.  Such perturbations have $\delta S \not = 0$.  Since all four of these eigenvalues are negative, any perturbations that change the value of $S$ will die out exponentially in time.  

This completes our analysis of the stability of the $S = 0$ equilibria in which all order parameters have length $r_0$.  The equilibria are stable with respect to an $M+2$ dimensional space of perturbations and neutrally stable with respect to an $M-2$ dimensional space of perturbations.  The neutrally stable perturbations all maintain the condition $S=0$, so this space of equilibria is a stable set of solutions.

\subsubsection{Equilibria with $S = 0$ and one or more incoherent groups}
Suppose, in an equilibrium with $S = 0$, that $m$ of the groups (denoted $\sigma = 1, 2, ..., m$) are incoherent, so that $\alpha_1 = \alpha_2 = \dots = \alpha_m = 0$,  and the rest of the groups are coherent, with $|\alpha_{m+1}| = |\alpha_{m+2}| = \dots = |\alpha_M| = r_0$ and $\sum_{\sigma = m+1}^M \alpha_\sigma = 0$.  We first consider the stability of this equilibrium for the case where $m \ge 2$.  We perturb the equilibrium in such a way that $\alpha_1 = -\alpha_2$ (where $\alpha_1$ and $\alpha_2$ are small but non-zero) and all other $\alpha_\sigma$ are unchanged.  This perturbation leaves $S=0$.  Linearizing Eq.\ \eqref{eq: lowdim3}, we find that
\begin{equation}
\label{eq: incoherent}
\frac{d\alpha_{1,2}}{dt} = \left(\frac{K}{2}(1+b)-1\right) \alpha_{1,2}.
\end{equation}
For $K > K_c$, this yields exponential growth for $\alpha_1$ and $\alpha_2$ [preserving the initial requirement that $\alpha_1 (t) = -\alpha_2 (t)$] and hence instability for the equilibrium.

It remains to consider the stability of $S=0$ equilibria in which only one group is incoherent.  This is discussed in Appendix \ref{app: demon2}.

\subsubsection{Equilibria with $S\not = 0$}
Since all the order parameters, as well as $S$, can be assumed to lie on the real line for equilibria with $S \not = 0$, we have $S^* = S$ and $r_\sigma^* = r_\sigma$ for all $\sigma$.  We will consider strictly imaginary perturbations $\delta r_\sigma$ to these equilibria and show that they are unstable given those perturbations. Then if we define $\delta S = \sum_{\sigma=1}^M \delta r_\sigma$, we also have $\delta S^* = -\delta S$ and $\delta r_\sigma^* = - \delta r_\sigma$.  Considering only the first-order differential terms of Eq.\ \eqref{eq: lowdim3} and incorporating the above assumptions, we obtain
\begin{equation}
\label{eq: lowdimsnotzero}
\frac{d \delta r_\sigma}{d t} = -\left[1 + \frac{K}{2} \left((1+b)(r_\sigma^2-1) - 2 r_\sigma b S\right)\right] \delta r_\sigma - \frac{Kb}{2} (1 + r_\sigma^2) \delta S.
\end{equation}
Taking into account the equilibrium condition of Eq.\ \eqref{eq: snonzero}, this equation becomes
\begin{equation}
\label{eq: lowdimsnotzero2}
\frac{2}{Kb}\frac{d \delta r_\sigma}{d t} = (r_\sigma + \frac{1}{r_\sigma})\left[S \delta r_\sigma - r_\sigma \delta S\right].
\end{equation}
Now we rewrite this as a vector equation:
\begin{equation}
\label{eq: snonzerovec}
\frac{2}{Kb}\frac{d \vec{\delta r}}{d t} = (D + D^{-1})(S\mathds{1} - \hat{\mathds{1}}D)\vec{\delta r},
\end{equation}
where $\hat{\mathds{1}}$ is the $M\times M$ matrix with all entries equal to one, and $D$ is the diagonal matrix $D = \operatorname{diag}[r_1, r_2, ..., r_M]$.  We also define the matrices $E = \operatorname{diag}[\sqrt{1+r^{-2}_1}, \sqrt{1+r_2^{-2}}, ..., \sqrt{1+r_M^{-2}}]$
and $W = S(D + D^{-1}) - E^2 D \hat{\mathds{1}} D$.  Equation \eqref{eq: snonzerovec} can then be written as
\begin{equation}
\label{eq: snonzerovec2}
\frac{2}{K b} \frac{d\vec{\delta \alpha}}{dt} = W \vec{\delta \alpha}.
\end{equation}
If we assume an exponential solution for $\vec{\delta r}$, then we see from Eq.\ \eqref{eq: snonzerovec2} that the equilibrium will be unstable if the matrix $W$ has any positive eigenvalues.  The eigenvalues of $W$ are the same as those of the matrix $\hat W = E^{-1}W E$, which can be written as 
\begin{equation}
\label{eq: mhat}
\hat W = S(D + D^{-1}) - (ED)\hat{\mathds{1}} (ED).
\end{equation}
The matrix $\hat W$ is symmetric and has real entries, so the largest eigenvalue $\lambda_{max}$ of $\hat W$ must satisfy
\begin{equation}
\label{eq: lambdamax}
\lambda_{max} \ge \frac{\vec u^T \hat W \vec u}{\vec u^T \vec u},
\end{equation}
for any real vector $\vec{u}$.  We choose a vector $\vec u_0$ such that $\hat{\mathds{1}}(ED)\vec u_0 = 0$.  This condition can be written as 
\begin{equation}
\label{eq: uconstraint}
\sum_{\sigma = 1}^M  u_0^\sigma \sqrt{r_\sigma^2 + 1} = 0,
\end{equation}
where $u_0^\sigma$ is the $\sigma$th component of the vector $\vec u_0$.  This represents a one-dimensional constraint on $\vec u_0$, so there exists an $M-1$ dimensional subspace of valid $\vec u_0$ vectors.  It now remains to show that we can find a $\vec u_0$ for which the right hand side of Eq.\ \eqref{eq: lambdamax} is positive.  To accomplish this, we note that for any $\sigma \in \{1,2,...,M\}$, Eq.\ \eqref{eq: snonzero} can be solved for $S$, giving
\begin{equation}
\label{eq: S}
S = \frac{r_\sigma (1+b)}{b} \left(\frac{r_0^2 - r_\sigma^2}{1-r_\sigma^2}\right),
\end{equation}
where $r_0^2$ is defined as in Eq.\ \eqref{eq: rsquared}.  In order to satisfy the assumption that $S > 0$, we must have that $r_\sigma^2 <  r_0^2$ if $r_\sigma > 0$ and $r_\sigma^2 > r_0^2$ if $r_\sigma < 0$.  This implies that the magnitude of the negative root of Eq.\ \eqref{eq: snonzero} is always greater than either of the two positive roots.  Thus we cannot have $S > 0$ unless at least two of the values of $r_\sigma$ are positive; if all but one of the $r_\sigma$ values were negative, the sum of the $r_\sigma$ values would have to be negative.  

Let $r_{\sigma_1}$ and $r_{\sigma_2}$ be any two positive order parameters.  Choose the vector $\vec u_0$ so that all of the components $u_0^\sigma$ are zero when $\sigma \neq \sigma_1$ or $\sigma_2$, and choose $u_0^{\sigma_1}$ and $u_0^{\sigma_2}$ so that Eq.\ \eqref{eq: uconstraint} is satisfied.  Then Eq.\ \eqref{eq: lambdamax} gives
\begin{equation}
\label{eq: lambdamax2}
\lambda_{max} \ge \frac{(u_0^{\sigma_1})^2 (r_{\sigma_1} + r_{\sigma_1}^{-1}) + (u_0^{\sigma_2})^2 (r_{\sigma_2} + r_{\sigma_2}^{-1})}{(u_0^{\sigma_1})^2 + (u_0^{\sigma_2})^2},
\end{equation}
which is positive.  Thus, this choice for $\vec u_0$ shows that $\lambda_{max} > 0$.  The matrix $W$ therefore has a positive eigenvalue, implying that there exist perturbations $\vec{\delta \alpha}$ that grow exponentially in time.  Consequently, any member of this class of equilibria (those with $S \neq 0$) is unstable.  Thus the above analysis confirms our numerical observation that, among all possible equilibria, the only one that is not unstable is the one with all order parameter magnitudes equal to $r_0$ and $S = 0$.

\section{Nonidentical Groups}
\label{sec: nonidenticalGroups}
\subsection{Formulation}
\label{subsec: nonidentformulation}
In order to make our model more realistic and reconcile the low-dimensional analysis with numerical simulations of the full system reported in Sec.\ \ref{subsec: simulations}, we introduce small deviations in the frequency distribution centers $\Delta_\sigma$and widths $\bar{\omega}_\sigma$, as well as in the coupling strengths $K_{\sigma \sigma'}$ of the oscillator groups.  We let $\epsilon$ be a small positive number and let $\bar{\omega}_\sigma = \epsilon \omega_\sigma$ be the average natural frequency of the oscillators in the $\sigma$th group.  We also let $\epsilon \delta_\sigma$ be the deviation in the width of the frequency distribution of group $\sigma$, and $\epsilon k_{\sigma \sigma'}$ be the deviation in coupling strength between groups $\sigma$ and $\sigma'$.  Thus
\begin{equation}
\label{eq: deltadeviation}
\Delta_\sigma = 1 + \epsilon \delta_\sigma
\end{equation}
and
\begin{equation}
\label{kdeviation}
K_{\sigma \sigma'} = \left\{
     \begin{array}{lr}
       K + \epsilon k_{\sigma \sigma'} & : \sigma = \sigma'\\
       -K b + \epsilon k_{\sigma \sigma'} & : \sigma \not = \sigma'\\
     \end{array}
   \right. .
\end{equation}
Letting $\tau = \epsilon t$ be a long characteristic timescale for the system, we then assume a configuration in which $S=0$ and $\alpha_\sigma (t, \tau) = r_0 e^{i \psi_\sigma (\tau)} (1+ \epsilon \alpha^{(1)}_\sigma (t, \tau))$, where $\epsilon \alpha^{(1)}_\sigma (t, \tau)$ represents a small, ``fast" perturbation in $\alpha_\sigma$, and the angles $\psi_\sigma$, by virtue of their assumed dependence on $\tau$, are allowed to change slowly in time.  Substituting this ansatz for $\alpha_\sigma$ into Eq.\ \eqref{eq: lowdim3} and collecting terms that are first order in $\epsilon$, we get
\begin{eqnarray}
\label{eq: nonidentical}
\frac{\partial(\alpha_\sigma^{(1)})}{\partial t} + \alpha_\sigma^{(1)} + \frac{K}{2}\left[(1+b)\left[r_0^2 \alpha_\sigma^{(1)*} + (2 r_0^2-1) \alpha_\sigma^{(1)}\right] - b r_0 e^{i \psi_\sigma} S^{(1)*} + \frac{b}{r_0} e^{-i \psi_\sigma} S^{(1)}\right]  \nonumber \\
= -\delta_\sigma -i\left(\frac{d\psi_\sigma}{d\tau} - \omega_\sigma\right) - \frac{1}{2} \sum_{\sigma' = 1}^M k_{\sigma \sigma'} (r_0^2 e^{i(\psi_{\sigma} - \psi_{\sigma'})} - e^{-i(\psi_{\sigma} - \psi_{\sigma'})}),
\end{eqnarray}
where $S^{(1)} = \sum_{\sigma = 1}^M r_0 e^{i \psi_\sigma} \alpha_\sigma^{(1)}$.  Note the similarity to Eq.\ \eqref{eq: lowdimpert}.  Just as in our stability analysis in Sec.\ \ref{subsubsec: szero}, we can combine this system of coupled ODEs and their conjugates into a vector equation:
\begin{equation}
\label{eq: nonidenticaldouble}
\frac{d}{dt}
\left( \begin{array}{c}
\vec{\alpha^{(1)}} \\
\vec{\alpha^{(1)}}^* \end{array} \right) - 
A \left( \begin{array}{c}
\vec{\alpha^{(1)}} \\
\vec{\alpha^{(1)}}^* \end{array} \right) = 
\left( \begin{array}{c}
\vec{Q}\\
\vec{Q^*} \end{array} \right),
\end{equation}
where $A$ is defined as in Eq.\ \eqref{eq: AandB},
\begin{equation}
\label{eq: alphadef}
\vec{\alpha^{(1)}} = \left(\begin{array}{c}
\alpha_1^{(1)}\\
\alpha_2^{(1)}\\
\vdots\\
\alpha_M^{(1)} \end{array} \right),
\end{equation}
and the components of $\vec{Q}$ are given by
\begin{equation}
\label{eq: Q}
Q_\sigma = -\delta_\sigma -i\left(\frac{d\psi_\sigma}{d\tau} - \omega_\sigma\right) - \frac{1}{2} \sum_{\sigma' = 1}^M k_{\sigma \sigma'} (r_0^2 e^{i(\psi_{\sigma} - \psi_{\sigma'})} - e^{-i(\psi_{\sigma} - \psi_{\sigma'})}).
\end{equation}
We can find constraints on the value of $\vec Q$ by multiplying Eq.\ \eqref{eq: nonidenticaldouble} on the left by $\left( \begin{array}{ccc} \vec u^T & \Big | & -\vec u^T \end{array} \right)$, where $\vec u$ is a vector satisfying $\vec \xi^T \vec u = 0$ as in Sec.\ \ref{subsubsec: szero}.  The vector $\left( \begin{array}{ccc} \vec u^T & \Big | & -\vec u^T \end{array} \right)$ is a left eigenvector of $A$ with eigenvalue 0, so the second term in Eq.\ \eqref{eq: nonidenticaldouble} vanishes, leaving
\begin{equation}
\label{eq: nonidenticalu}
\left( \begin{array}{ccc} \vec u^T & \Big | & -\vec u^T \end{array} \right) \frac{d}{dt}
\left( \begin{array}{c}
\vec{\alpha^{(1)}} \\
\vec{\alpha^{(1)}}^* \end{array} \right) = \left( \begin{array}{ccc} \vec u^T & \Big | & -\vec u^T \end{array} \right) \left( \begin{array}{c}
\vec{Q}\\
\vec{Q^*} \end{array} \right).
\end{equation}
The right-hand side of Eq.\ \eqref{eq: nonidenticalu} is effectively constant on the $t$ timescale.  If it is non-zero, then $\vec{\alpha^{(1)}}$ will diverge linearly with $t$ on the fast timescale.  We prevent this by demanding that 
\begin{equation}
\label{eq: nonidenticalu2}
\left( \begin{array}{ccc} \vec u^T & \Big | & -\vec u^T \end{array} \right) \left( \begin{array}{c}
\vec{Q}\\
\vec{Q^*} \end{array} \right) = 0,
\end{equation}
or $\vec u^T (\operatorname{Im} \vec{Q}) = 0$, where $\operatorname{Im} \vec{Q}$ denotes the imaginary part of $\vec{Q}$.  This equation holds for all valid choices of $\vec u$ if and only if $\operatorname{Im} \vec{Q}$ is in the space spanned by the vectors $\vec c$ and $\vec s$ (see Eq.\ \eqref{eq: cands}).  Thus, at each time $\tau$ there are scalars $X$ and $Y$ such that 
\begin{equation}
\label{eq: psiodes}
\frac{d\psi_\sigma}{d\tau} = \omega_\sigma - \frac{1}{2} \sum_{\sigma' = 1}^M \left[ k_{\sigma \sigma'}(r_0^2 + 1)\sin(\psi_{\sigma} - \psi_{\sigma'})\right] + X \cos \psi_\sigma + Y \sin \psi_\sigma.
\end{equation}
We find the coefficients $X$ and $Y$ by recalling that $S=0$ and therefore
\begin{equation}
\label{eq: cosinessineszero}
\sum_{\sigma = 1}^M \cos{\psi_\sigma} = \sum_{\sigma = 1}^M \sin{\psi_\sigma} = 0.
\end{equation}
Taking the $\tau$ derivative of this equation gives
\begin{equation}
\label{eq: cosinessineszeroderiv}
\sum_{\sigma = 1}^M \sin{\psi_\sigma}\frac{d\psi_\sigma}{d\tau} = \sum_{\sigma = 1}^M \cos{\psi_\sigma}\frac{d\psi_\sigma}{d\tau} = 0.
\end{equation}
Substituting Eq.\ \eqref{eq: psiodes} into Eq.\ \eqref{eq: cosinessineszeroderiv}, we obtain a system of two linear equations that uniquely specify $X$ and $Y$ at each time $\tau$:
\begin{equation}
\label{eq: cosinessineszeroderiv2}
\left\{
     \begin{array}{l}
       \begin{array}{l}\left(\sum_{\sigma = 1}^M \cos^2 \psi_\sigma\right) X + \left(\sum_{\sigma = 1}^M \sin \psi_\sigma \cos \psi_\sigma\right) Y \\ \hspace*{2 cm}= -\sum_{\sigma = 1}^M \omega_\sigma \cos \psi_\sigma + \frac{1}{2} \sum_{\sigma = 1}^M \sum_{\sigma' = 1}^M k_{\sigma \sigma'}(r_0^2 + 1)\sin(\psi_{\sigma} - \psi_{\sigma'})\cos \psi_\sigma \end{array}\\
       \begin{array}{l}\left(\sum_{\sigma = 1}^M \sin \psi_\sigma \cos \psi_\sigma\right) X + \left(\sum_{\sigma = 1}^M \sin^2 \psi_\sigma\right) Y \\  \hspace*{2 cm}= -\sum_{\sigma = 1}^M \omega_\sigma \sin \psi_\sigma + \frac{1}{2} \sum_{\sigma = 1}^M \sum_{\sigma' = 1}^M k_{\sigma \sigma'}(r_0^2 + 1)\sin(\psi_{\sigma} - \psi_{\sigma'})\sin \psi_\sigma\end{array}\\
       \end{array}
   \right. .
\end{equation}
Solving Eq.\ \eqref{eq: cosinessineszeroderiv2}  gives $X$ and $Y$ as functions of the angles $\psi_\sigma$.  Insertion of these functions into Eq.\ \eqref{eq: psiodes} then gives the desired slow timescale evolution equation for $\vec \psi$.  Note that, because the values of $\delta_\sigma$ are real, they do not appear in Eq.\ \eqref{eq: cosinessineszeroderiv2} and hence do not contribute to the $O(\epsilon)$ slow timescale evolution.  Also note that $\sum \cos \psi_\sigma$ and $\sum \sin \psi_\sigma$ are constants of the motion for this system (see Eq.\ \eqref{eq: cosinessineszeroderiv}).

The two linear algebraic equations in Eq. \eqref{eq: cosinessineszeroderiv2} are insoluble when the associated determinant,
\begin{equation}
\label{eq: T}
T \equiv \left(\sum_{\sigma} \cos^2{\psi_\sigma}\right) \left(\sum_{\sigma} \sin^2{\psi_\sigma}\right) - \left(\sum_\sigma \sin{\psi_\sigma} \cos{\psi_\sigma}\right)^2 = (\vec{c}^T \vec{c})(\vec{s}^T \vec{s}) - (\vec{c}^T \vec{s})^2,
\end{equation}
is equal to zero.  This happens when $\vec{c}$ and $\vec{s}$ (the real and imaginary parts of $\vec{\xi}$) are parallel, implying $\tan{\psi_1} = \tan{\psi_2} = \hdots = \tan{\psi_M}$.  Thus $T = 0$ occurs when all the $\psi_\sigma$ are at two angles separated by $\pi$ (i.e., the order parameters lie on a single line through the origin).  Because we still need $r_\sigma \approx r_0$ and $S \approx 0$, $T = 0$ can only occur when $M$ is an even number.  For $M = 2$, $T$ is always zero ($\psi_1 = \psi_2 + \pi$).  For $M = 4$ the four order parameters occur as two $\pi$-separated pairs, so that $T = 0$ whenever the two pairs coincide (Sec.\ \ref{m3and4}).  For $M$ even, as $M$ increases beyond $4$, it becomes unlikely that $T = 0$ will ever occur.  Our short/fast timescale expansion assumes that $T = O(\epsilon^0)$.  Thus, in situations where $T$ approaches zero, our expansion breaks down (when $T$ becomes $O(\epsilon)$).  In Sec.\ \ref{nonidentnumerical} we will discuss the possible occurrence and implications of $T$ becoming small in the context of numerical experiments.

\subsection{The Examples of $M=3$ and $M=4$}
\label{m3and4}
When $M=3$, the space of vectors $\vec u$ satisfying $\vec \xi^T \vec u = 0$ is one dimensional, spanned by the vector
\begin{equation}
\label{eq: u0}
\vec{u_0} = \left( \begin{array}{c} 1\\1\\1\end{array}\right).
\end{equation}
In this case, Eq.\ \eqref{eq: nonidenticalu2} gives 
\begin{equation}
\label{eq: m3ode}
\sum_{\sigma =1}^3 \left[\frac{\partial \psi_\sigma}{\partial \tau} - \omega_\sigma + \frac{1}{2}(r_0^2+1)\sum_{\sigma'=1}^3 \left[k_{\sigma \sigma'} \sin(\psi_{\sigma}-\psi_{\sigma'})\right]\right] = 0.
\end{equation}
Because we must have $S=0$ at each time step, the phases $\psi_\sigma$ must be separated by $2\pi/3$, and all three phases must change at the same rate.  If we assume $\psi_2 - \psi_1 = \psi_3 - \psi_2 = 2\pi/3$, Eq.\ \eqref{eq: m3ode} becomes
\begin{equation}
\label{eq: m3ode2}
\frac{\partial \psi_\sigma}{\partial \tau} = w_\omega + w_k,
\end{equation}
where
\begin{equation}
\label{eq: wwwk}
 w_\omega = \frac{\omega_1 + \omega_2 + \omega_3}{3}, w_k = \frac{\sqrt{3}}{12}(r_0^2+1)\left[k_{12} +k_{23} +k_{31} -k_{21} -k_{32} -k_{13}\right],
 \end{equation}
and $\sigma = 1, 2, $ or $3$.  We thus see that the phases $\psi_\sigma$ change at the same constant rate given by the right-hand side of Eq.\ \eqref{eq: m3ode2}.  Note that, if the perturbed coupling matrix is symmetric, $k_{\sigma \sigma'} = k_{\sigma' \sigma}$, then $w_k = 0$, and the rotation is at the average frequency $w_\omega$.

In the $M=4$ case, $S = 0$ is obtained by having two pairs of order parameters whose phases are separated by $\pi$.  We therefore assume without loss of generality that $\psi_3 = \psi_1 + \pi$ and $\psi_4 = \psi_2 + \pi$.  We can then express the vectors $\vec u$ independently of the angles $\psi_\sigma$.  One possible choice is 
\begin{equation}
\label{eq: uvectors4}
\vec u_1 = \left(\begin{array}{c}
1\\0\\1\\0\end{array}\right)
\mbox{   and    }
\vec u_2 = \left(\begin{array}{c}
0\\1\\0\\1\end{array}\right).
\end{equation}
It is clear that these vectors satisfy $\vec \xi^T \vec u =0$ given the above conditions on the phases $\psi_\sigma$.  If we use vector $\vec u_1$ in Eq.\ \eqref{eq: nonidenticalu2}, we obtain
\begin{equation}
\label{eq: u1ode}
\frac{\partial \psi_1}{\partial \tau} - \omega_1 + \frac{1}{2}\sum_{\sigma=1}^M k_{1 \sigma} (r_0^2+1)\sin(\psi_1-\psi_\sigma) +\frac{\partial \psi_3}{\partial \tau} - \omega_3 + \frac{1}{2}\sum_{\sigma=1}^M k_{3 \sigma} (r_0^2+1)\sin(\psi_3-\psi_\sigma)=0.
\end{equation}
In order to preserve $S=0$, it is necessary that $\psi_1$ and $\psi_3$ change at the same rate.  Letting $\partial \psi_3/\partial \tau = \partial \psi_1/\partial \tau$ in Eq.\ \eqref{eq: u1ode} and simplifying yields
\begin{equation}
\label{eq: u1ode2}
\frac{\partial \psi_1}{\partial \tau} - \frac{\omega_1 + \omega_3}{2} + \frac{1}{4}(r_0^2+1)(k_{12}-k_{32} - k_{14} + k_{34})\sin(\psi_1-\psi_2)=0.
\end{equation}
Using vector $\vec u_2$ in Eq.\ \eqref{eq: nonidenticalu2} gives a similar ODE for $\psi_2$:
\begin{equation}
\label{eq: u2ode}
\frac{\partial \psi_2}{\partial \tau} - \frac{\omega_2 + \omega_4}{2} + \frac{1}{4}(r_0^2+1)(k_{21}-k_{41} - k_{23} + k_{43})\sin(\psi_2-\psi_1)=0.
\end{equation}
If we subtract Eq.\ \eqref{eq: u2ode} from Eq.\ \eqref{eq: u1ode2} and define $\Delta \psi = \psi_1-\psi_2$, we obtain an ODE for the evolution of $\Delta \psi$:
\begin{equation}
\label{eq: deltapsiode}
\frac{\partial \Delta \psi}{\partial\tau} - \Omega + \mathcal{H} \sin \Delta \psi = 0,
\end{equation}
where $\Omega = (\omega_1 - \omega_2 + \omega_3 - \omega_4)/2$ and $\mathcal{H} = (r_0^2+1)(k_{12}+k_{21}+k_{34}+k_{43}-k_{14}-k_{41}-k_{23}-k_{32})/4$.
Once $\Delta \psi = (\psi_1 - \psi_2)$ is found from Eq.\ \eqref{eq: deltapsiode}, $\psi_1$ and $\psi_2$ can be determined by inserting $\Delta \psi$ into \eqref{eq: u1ode2} and \eqref{eq: u2ode}.  Note that $\Omega$ depends only on the frequency perturbations $\omega_\sigma$, while the term $\mathcal{H}$ responsible for interaction between groups depends only on the off-diagonal coupling perturbations, $k_{\sigma \sigma'}$ for $\sigma \neq \sigma'$.  The solution to Eq.\ \eqref{eq: deltapsiode} depends on the relative sizes of $\Omega$ and $\mathcal{H}$.  If $|\Omega| > |\mathcal{H}|$, the solution is periodic, and the time derivative of $\Delta \psi$ always has the same sign.  If $|\Omega| \le |\mathcal{H}|$, then $\Delta \psi$ eventually reaches a constant value, at which point all groups rotate at the same speed.

In this analysis of the $M = 4$ case we assume that the pairs remain the same, i.e., that groups $1$ and $3$ always remain paired.  However, it is possible for the groups to switch partners when the group order parameters overlap ($\Delta \psi = 0$ or $\pi$).  In this case the vectors $\vec{u_1}$ and $\vec{u_2}$ will change to reflect the new pairing.  This phenomenon will be discussed in Sec.\ \ref{nonidentnumerical}.

\subsection{Numerical Results}
\label{nonidentnumerical}
Returning to our numerical simulation of the low-dimensional system, we introduced the small perturbations $\omega_\sigma$, $\delta_\sigma$, and $k_{\sigma \sigma'}$, and integrated Eq.\ \eqref{eq: lowdim2}.  When these perturbations were included, the system no longer approached a steady-state equilibrium.  Rather, the group order parameters behaved as in the high-dimensional simulations of Sec.\ \ref{subsec: simulations}, establishing a configuration in which $S\approx 0$ and then evolving in the complex $\alpha$-plane in a way that kept $S$ near zero and $|\alpha_\sigma|$ nearly $r_0$ at all times. The similarity to the high-dimensional system indicates that the  behavior of the full system can be well understood by studying the low-dimensional case with nonidentical groups.  

\begin{figure}[H]
\centering
\includegraphics[width=16cm]{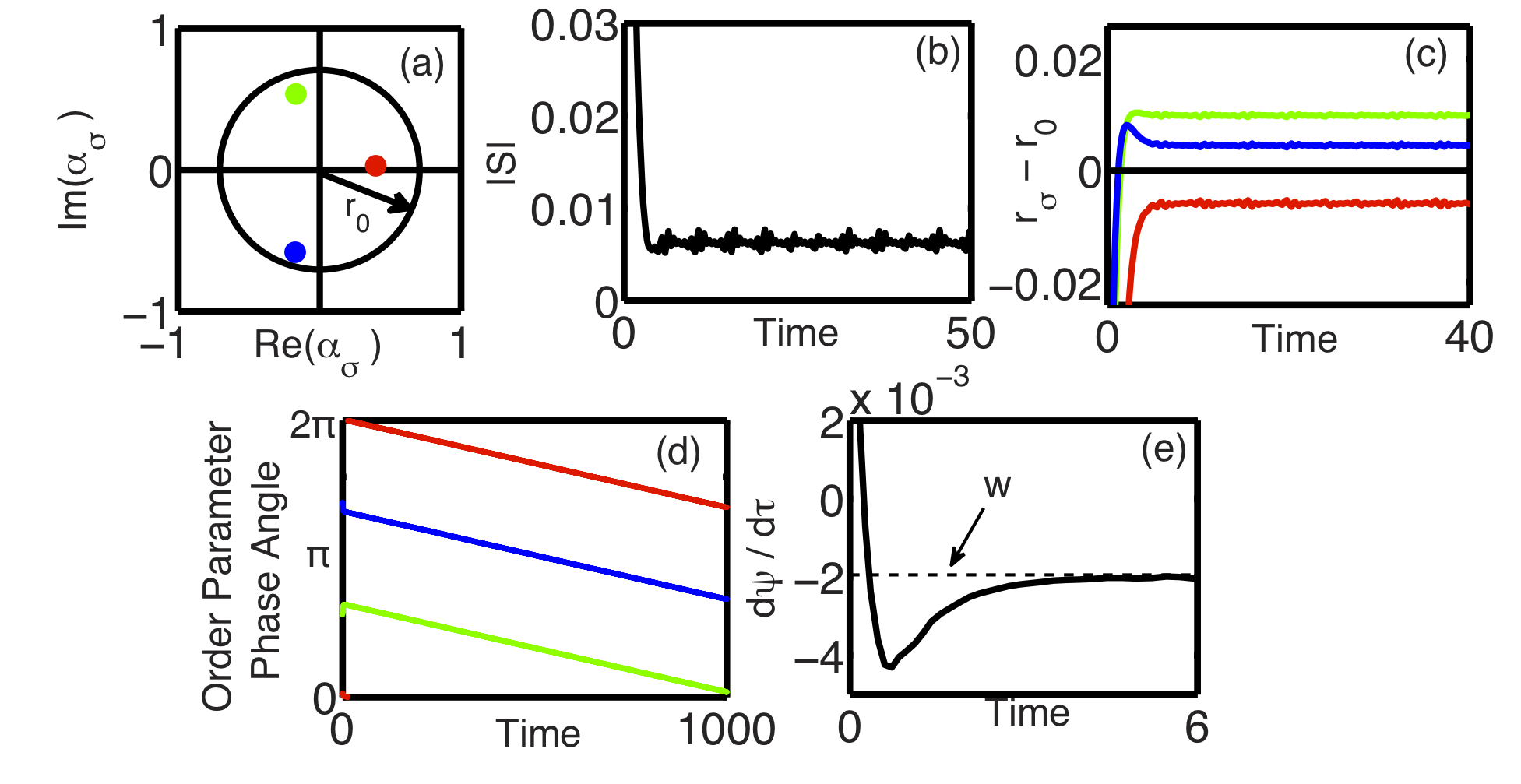}
\caption{\footnotesize{Results for a case with three groups ($M=3$), obtained by integrating Eq.\ \eqref{eq: lowdim2} with $\bar{\omega}_\sigma, \Delta_\sigma$, and $K_{\sigma \sigma'}$ defined as in Sec.\ \ref{subsec: nonidentformulation} and the perturbations $\omega_\sigma, \delta_\sigma$, and $k_{\sigma \sigma'}$ given by \eqref{eq: perts}.  (a) The initial conditions, $\alpha_1(0)$ (red dot), $\alpha_2(0)$ (green dot), $\alpha_3(0)$ (blue dot); (b) the evolution of $|S|$; (c) evolution of $r_\sigma$ with respect to $r_0$; (d) evolution of $\psi_1$, $\psi_2$, and $\psi_3$ (in radians); (e) evolution of $d\bar{\psi}/d\tau$, where $\bar{\psi} = (\psi_1 + \psi_2 + \psi_3)/3$.  In (e), the horizontal line denotes the value $w = w_\omega + w_k$ given in Eq.\ \eqref{eq: wwwk}.}}
\label{fig: Figure5}
\end{figure}

Figure \ref{fig: Figure5} shows a numerical example for a case with three groups ($M=3$).  In this example the parameter perturbations from exact symmetry are
\begin{equation}
\label{eq: perts}
\begin{array}{c} \epsilon\omega_1 =  0.0158, \ \epsilon\omega_2 = 0.0060, \ \epsilon\omega_3 =  -0.0033,\\
\ 
 \\ \epsilon\delta_1 = .0252, \ \epsilon\delta_2 = -0.0432, \ \epsilon\delta_3 = -0.0188,  \\
 \ 
 \\ \epsilon k = \left( \begin{array}{ccc}  0.0321  &  0.0034 & 0.0148\\
    0.0270 & -0.0224 & 0.0180\\
   -0.0090 &  0.0080 & -0.0306\end{array} \right). \end{array}
   \end{equation}
The initial conditions (Fig.\ \ref{fig: Figure5}(a)) are such that $S$ is not initially zero.  However, as time increases, $S$ is observed to move quickly towards zero (Fig.\ \ref{fig: Figure5}(b)), but due to the $O(\epsilon)$ parameter perturbations from symmetry, its magnitude, although small, does not become exactly zero.  Also, the order parameter magnitudes $r_\sigma$ approach $r_0$ given by Eq.\ \eqref{eq: rsquared} (Fig.\ \ref{fig: Figure5}(c)), but with some $O(\epsilon)$  deviation.  Figure \ref{fig: Figure5}(d) shows the time evolution of the order parameter phases $\psi_1, \psi_2$, and $\psi_3$, which, as predicted in Sec.\ \ref{m3and4}, approach an approximately evenly spaced, uniformly rotating configuration.  Figure \ref{fig: Figure5}(e) shows the time evolution of the average rotation rate $d\bar{\psi}/d\tau$, where $\bar{\psi} = (\psi_1+\psi_2+\psi_3)/3$, showing that $d\bar{\psi}/d\tau$ approaches the theoretical prediction $w_\omega + w_k$ given in Eq.\ \eqref{eq: m3ode2}.

\begin{figure}[h!]
\centering
\includegraphics[width=17cm]{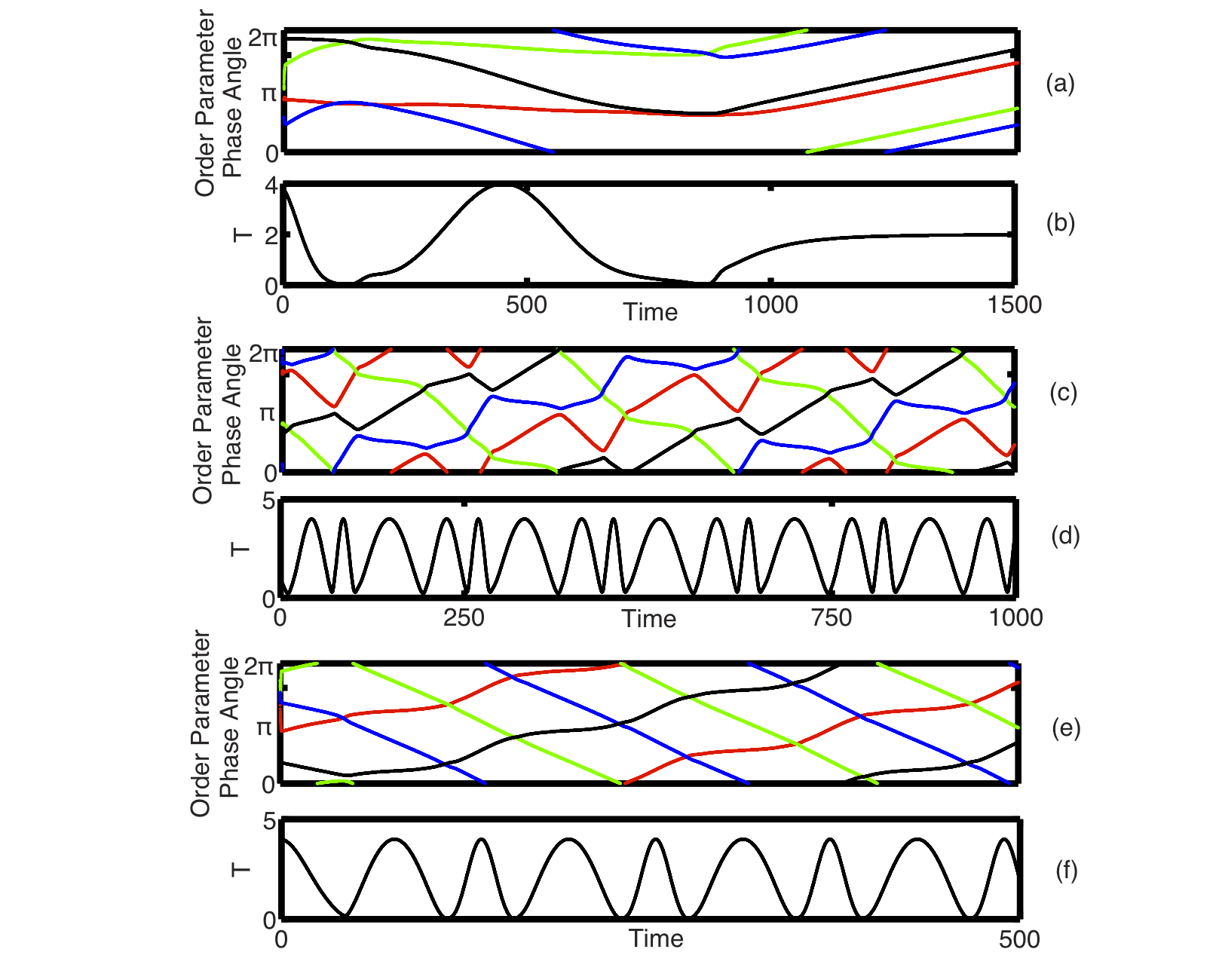}
\caption{\footnotesize{Numerical results for three cases with four groups ($M = 4$).  (a, c, e) The evolution of $\psi_1$, $\psi_2$, $\psi_3$, and $\psi_4$ (in radians); (b, d, f) the corresponding evolution of $T$.  The pair switches in each simulation are possible only when the value of $T$ nears zero.  In (a) and (b), this occurs at $t \approx 130$ and $t \approx 860$.  In (c) and (d), the pairs switch in a periodic manner throughout the simulation.  In (e) and (f), we see that $T \approx 0$ several times in the simulation but only one pair switch occurs, at $t \approx 35$.}}
\label{fig: Figure6}
\end{figure}

When $M=4$, numerical simulations exhibited both types of behavior predicted by Eq.\ \eqref{eq: deltapsiode}: when the values of $\omega_\sigma$ and $k_{\sigma \sigma'}$ were such that $|\Omega| > |\mathcal{H}|$, $\Delta \psi$ evolved periodically, and when $|\Omega| \le |\mathcal{H}|$, $\Delta \psi$ approached a fixed value.  However, the behavior was often more complicated than this, due to a pair switching phenomenon.  If, for example, the initial configuration was such that $\psi_3 = \psi_1 + \pi$ and $\psi_4 = \psi_2 + \pi$, it was sometimes possible for the groups to switch partners midway through the simulation (see Fig.\ \ref{fig: Figure6}(a) at $t \approx 130$ and $t \approx 860$), so that $\psi_1$ became paired with $\psi_2$ ($\psi_2 = \psi_1 + \pi$) and $\psi_3$ became paired with $\psi_4$ ($\psi_3 = \psi_4 + \pi$).  In Fig.\ \ref{fig: Figure6}(a), for example, when $t \lesssim 130$, the two pairings are red with black and blue with green; when $130 \lesssim t \lesssim 860$, the pairings are red with green and black with blue; and when $t \gtrsim 860$ the pairings are red with blue and black with green.  The values of $\Omega$ and $\mathcal{H}$ depend on the pairings of the groups, so it is possible to have $|\Omega| > |\mathcal{H}|$ before a pair switch and $|\Omega| \le |\mathcal{H}|$ afterwards.  This is what happens at $t \approx 860$ in Fig.\ \ref{fig: Figure6}(a).  Before $t \approx 860$, $|\Omega| > |\mathcal{H}|$.  After $t \approx 860$, $|\mathcal{H}| > |\Omega|$, and, for $t \gtrsim 860$, $\Delta \psi$ consequently approaches a fixed value, after which the whole configuration rotates rigidly.  Both the value of $\Delta \psi$ and the value of the rotation rate after $t \approx 860$ are correctly predicted by the fixed point of Eq.\ \eqref{eq: deltapsiode}.

The pair switching phenomenon is related to the failure of Eq.\ \eqref{eq: cosinessineszeroderiv2} to completely determine the system's behavior.  Figure \ref{fig: Figure6}(b) shows the value of the quantity $T$ defined in Eq.\ \eqref{eq: T} as a function of time.  We observe that $T \approx 0$ each time a pair switch occurs.  (For $t \gtrsim 860$ in Fig.\ \ref{fig: Figure6}(a, b) the rigid rotation of the configurations implies that $T$ becomes constant and that no pair switchings can occur after $t \approx 860$.)

Another $M = 4$ example is shown in Figs.\ \ref{fig: Figure6}(c, d).  In this case we observe that the switching assumes an apparently persistent time-periodic pattern.  For example, considering the group plotted in green in Fig.\ \ref{fig: Figure6}(c), we see that it is initially paired with red, then switches to being paired with blue, then with black, then back to red, etc.  Correspondingly, Fig.\ \ref{fig: Figure6}(d) shows that $T$ becomes periodic in time.  We observe that the periodic patterns in Figs.\ \ref{fig: Figure6}(c, d) continue for the length of very long numerical runs.

Note that it is also possible to have $T \approx 0$ when the two pairs pass through one another without switching.  For example, in Fig.\ \ref{fig: Figure6}(e, f), $T$ approaches zero many times throughout the simulation, but only the first of those occurrences ($t \approx 35$) corresponds to a pair switch.

\begin{figure}[h!]
\centering
\includegraphics[width=17cm]{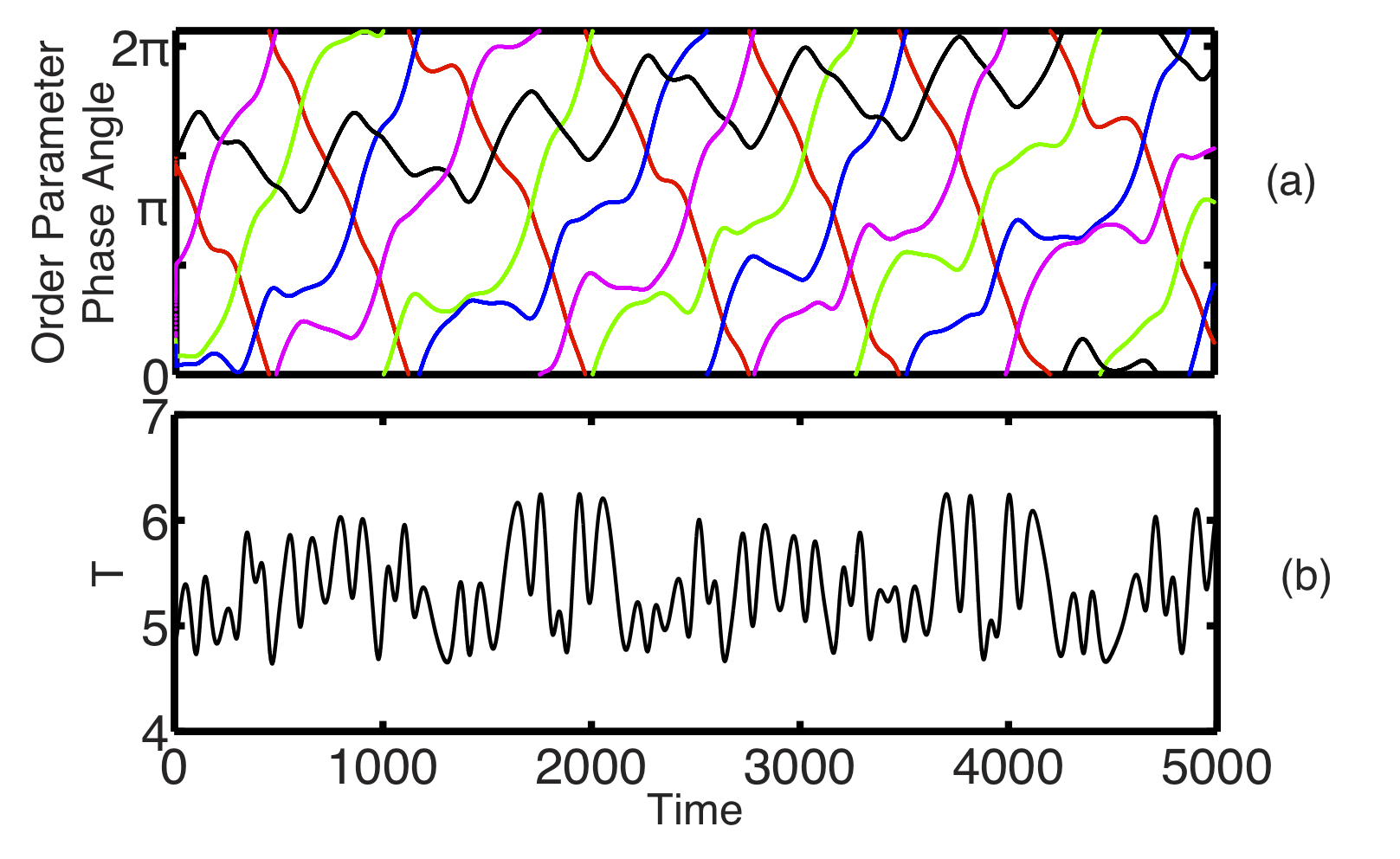}
\caption{\footnotesize{Numerical results for a case with five groups ($M = 5$).  (a) The evolution of $\psi_1$, $\psi_2$, $\psi_3$, $\psi_4$, and $\psi_5$ (in radians); (b) the corresponding evolution of $T$.  The value of $T$ remains large throughout the simulation, in contrast to the cases with $M = 4$ in Fig.\ \ref{fig: Figure6}.}}
\label{fig: Figure7}
\end{figure}

Figure \ref{fig: Figure7} shows results for a numerical simulation with $M=5$.  When $M=5$, it is impossible to have $T=0$ as long as the system maintains a configuration in which $S \approx 0$ and $r_\sigma \approx r_0$ for all $\sigma$.  Thus, the evolution of the phases $\psi_\sigma$ (Fig.\ \ref{fig: Figure7}(a)) always obeys Eq.\ \eqref{eq: psiodes}.  Note that the value of $T$ (Fig.\ \ref{fig: Figure7}(b)) is positive throughout the simulation.

\begin{figure}[h!]
\centering
\includegraphics[width=17cm]{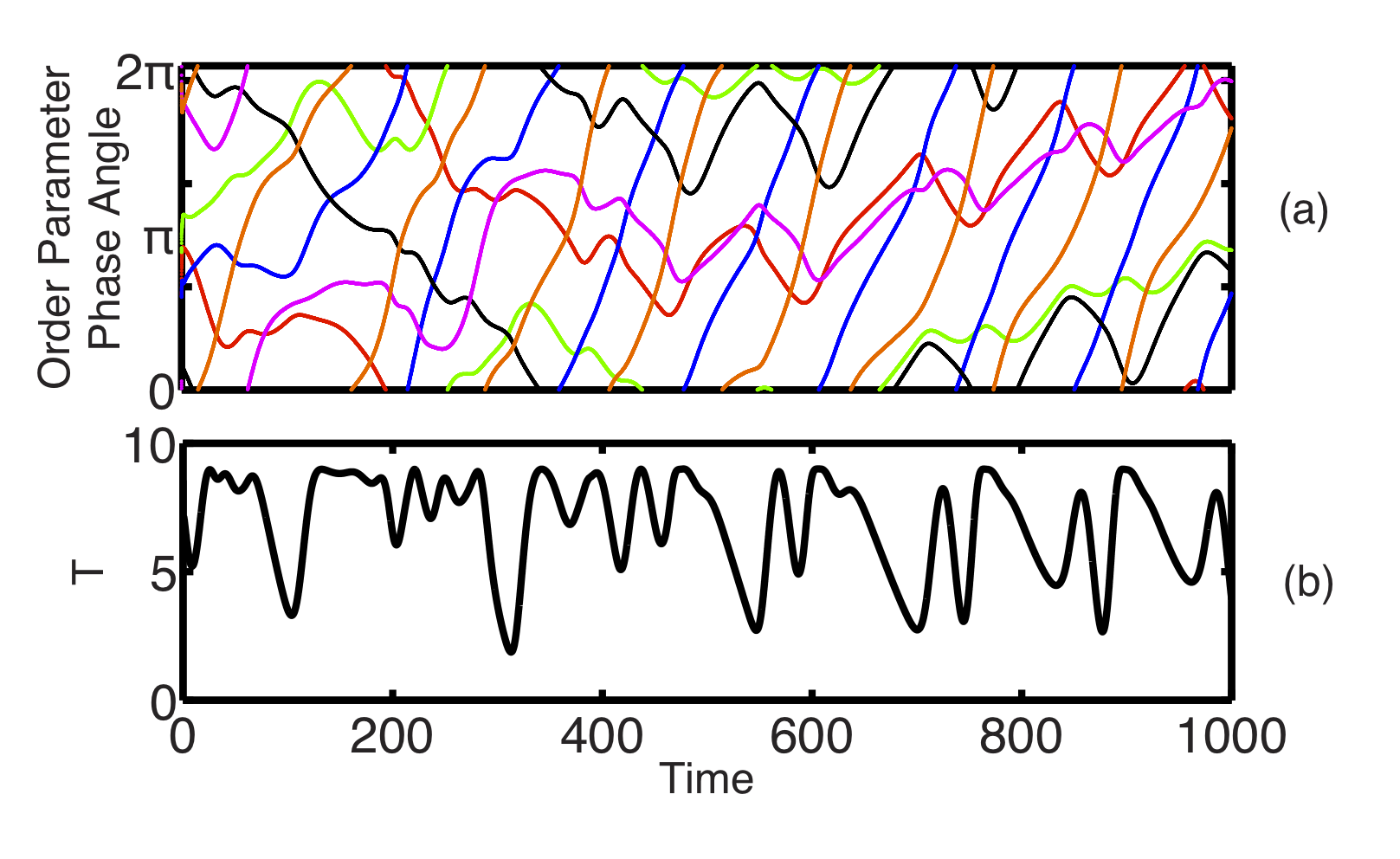}
\caption{\footnotesize{Numerical results for a case with six groups ($M = 6$).  (a) The evolution of $\psi_1$, $\psi_2$, $\psi_3$, $\psi_4$, $\psi_5$ and $\psi_6$ (in radians); (b) the corresponding evolution of $T$.}}
\label{fig: Figure8}
\end{figure}

The results of a simulation with $M=6$ can be seen in Fig.\ \ref{fig: Figure8}.  When $M=6$ it is possible, though unlikely, for $T=0$ to occur.  We observe that $T$ decreases significantly many times throughout the simulation, but it never approaches zero as closely as in the $M = 4$ case.  This means that Eq.\ \eqref{eq: cosinessineszeroderiv2} remains solvable for the duration of the simulation, so the dynamics of the phases can be described by Eq.\ \eqref{eq: psiodes}.  However, $T$ may sporadically become quite small in some $M = 6$ simulations, so Eq.\ \eqref{eq: psiodes} is not guaranteed to hold in all $M = 6$ cases.

Because of the difficulty in expressing equations of motion akin to Eqs.\ \eqref{eq: m3ode2} and \eqref{eq: deltapsiode} for $M = 5$ and $6$, it is not straightforward to characterize the possible behaviors of these systems.  For example, in some $M = 5$ and $M = 6$ simulations, the system exhibited periodic behavior.  In other simulations, the order parameters became locked in a static configuration, similar to the one seen in Fig.\ \ref{fig: Figure6}(a) after $t \approx 860$.

\section{Conclusion}
\label{sec:conclusion}

We have considered the dynamics of $M$ interacting groups of coupled oscillators, with ``attractive'' coupling within groups and ``repulsive'' coupling between groups. If the number of oscillators in each group is large, this system can be described using the low dimensional ansatz of Ott and Antonsen\cite{ref1}, which reduces the problem to a set of $M$ ODEs for the $M$ complex group order parameters.  Based on our analyses the typical dynamical evolution for such a system is as follows.  There is an initial phase of evolution in which members of the same group, which are attracted to each other, become synchronized and the group acquires a nonzero, complex order parameter.  In the symmetric case, where the group properties and interactions are all the same, the group order parameters relax on the same time scale to a manifold of neutrally stable equilibria in which all of the group order parameters have the same magnitude, and due to the repulsion between groups, the sum of the complex order parameters is zero.  Other equilibria also exist in this symmetric case.  However, we have shown them to be unstable.

In order to gain insight into the general non-symmetric case, we then introduced small perturbations to the group frequency distributions and coupling strengths. The effect of these small asymmetries is to introduce a slow time evolution to the phases of the group order parameters of the neutrally stable equilibria found in the symmetric case. To study this we used a slow/fast time scale analysis and obtained ODEs describing the slow-timescale motion of the order parameter phases.  In the $M=3$ case we observed that the group order parameters remained evenly spaced and rotated at a constant rate in the complex plane.  In the $M=4$ case the order parameters formed two pairs where members of the pairs differed in phase by $\pi$.  Depending on the asymmetry parameters the relative phases of the pairs either became locked or evolved periodically in time.  In the $M=5$ and higher cases the behavior of the order parameter phases became more complicated, showing a mixture of irregular and periodic motion.  Our conclusions above are supported by both analysis and numerical simulations of the low dimensional and high dimensional equations.

\section{Acknowledgements}
This work was supported by an NSF REU grant to the University of Maryland and by ONR grant N00014-07-1-0734.
Also thanks to user ghazwozza of the social media website reddit.com for contributing to the geometric ``tip-to-tail'' formulation of the $S = 0$ equilibria in Sec.\ \ref{subsec: szero}.
\label{sec:acknowledgements}

\appendix
\section{Demonstration that $B_\pm^2 \ge 4C_\pm$}
\label{app: demon}
To determine whether  $B_\pm^2 \ge 4C_\pm$, we take its derivative with respect to $P$ to obtain 
\begin{equation}
\label{pderiv}
\frac{d( B_\pm^2 - 4C_\pm)}{dP} = \pm K^2 b r_0^2\left[- (1 + b) + \frac{b}{2}(M \pm P r_0^2)\right].
\end{equation}
Note that the second derivative with respect to P is always positive, so all extrema of this function are minima.  If we choose the ``$-$'' option in Eq.\ \eqref{pderiv}, we find that this derivative is zero when
\begin{equation}
\label{eq: pextreme1}
P = r_0^{-2}(M - 2(1+b)/b).
\end{equation}
Then there are two cases depending on the value of $M$.  If $M \ge 2(1+b)/b$, then the value of  $B_-^2 - 4C_-$ is positive at the minimum, as hoped.  If $M < 2(1+b)/b$, then Eq.\ \eqref{eq: pextreme1} gives a negative value, but $P$ must be at least zero so the minimum value of $B_-^2 - 4C_-$ occurs outside the possible range of values for $P$.  Therefore, $B_-^2 - 4C_-$ is increasing on the interval $P \in [0, M]$ so we need only check its value at $P =0$ to ensure it is always positive.  Its value at $P = 0$ can be expressed as $((K(1+b)-2) - KMb/2)^2 + KMb(K(1+b)-2)(1-r_0^2)$, which is positive. 

Similarly, if we choose the ``$+$'' option in Eq.\ \eqref{pderiv}, then the derivative is zero when
\begin{equation}
\label{eq: pextreme2}
P = r_0^{-2}(2(1+b)/b - M).
\end{equation}
Again there are two cases depending on the value of $M$. If $M \ge (1+b)/b$, then the value of  $B_+^2 - 4C_+$ is nonnegative.  If $M < (1+b)/b$, then Eq.\ \eqref{eq: pextreme2} gives a value greater than $M$, so the minimum value for $B_+^2 - 4C_+$ again occurs outside the possible range of values for $P$.  Therefore, $B_+^2 - 4C_+$ is decreasing on the interval $P \in [0, M]$ so we need only check its value at $P = M$ to ensure it is always positive.  Its value at $P = M$ is $((K(1+b)-2) - KMb(1+r_0^2)/2)^2$ which is clearly positive as well. Thus $B_\pm^2 \ge 4C_\pm$.

\section{Instability of $S = 0$ equilibria with exactly one incoherent group}
\label{app: demon2}
Numerical simulations imply that equilibria in which $S=0$ and exactly one group is incoherent are unstable.  However, because the stability analysis of this case is algebraically intensive and the numerical evidence for instability is compelling, we prove instability only for the more tractable special case in which the nonzero order parameters are evenly spaced on a circle of radius $r_0$ in the complex plane.  Defining $\Delta \theta = 2\pi /(M-1)$ and letting $\delta \alpha_\sigma$ be small complex numbers, we suppose that
\begin{equation}
\label{eq: evenlyspaced}
\alpha_\sigma = e^{i \sigma \Delta \theta}(r_0 + \delta \alpha_\sigma)
\end{equation}
for $\sigma = 1, 2, ..., M-1$, and $\alpha_{M} = \delta \alpha_M$.  Substituting these quantities into Eq.\ \eqref{eq: lowdim3} and collecting terms that are first order in $\delta \alpha_\sigma$ gives
\begin{equation}
\label{eq: evenlyspacedpert1}
\frac{d(\delta \alpha_\sigma)}{dt} = -\frac{K}{2} r_0^2 (1+b) (\delta \alpha_\sigma + \delta \alpha_\sigma^*)  - \frac{K}{2} b\left(\delta S e^{-i \sigma \Delta \theta} - r_0^2 \delta S^* e^{i \sigma \Delta \theta}\right)
\end{equation}
for  $\sigma = 1, 2, ..., M-1$, and 
\begin{equation}
\label{eq: evenlyspacedpert2}
\frac{d(\delta \alpha_M)}{dt} = -\frac{K}{2} r_0^2 (1+b) \delta \alpha_M  - \frac{K}{2} b\delta S.
\end{equation}
In order to describe the collective behavior of the perturbed system, we define 
\begin{equation}
\label{eq: alphabar}
\bar{\alpha} = \sum_{\sigma = 1}^{M-1} e^{i \sigma \Delta \theta} \delta \alpha_\sigma \mbox{   and   }  \bar{\alpha}_* = \sum_{\sigma = 1}^{M-1} e^{i \sigma \Delta \theta} \delta \alpha_\sigma^*,
\end{equation}
and multiply Eq.\ \eqref{eq: evenlyspacedpert1} on both sides by $\exp(i \sigma \Delta \theta)$ and sum over all values of $\sigma$ from $1$ to $M-1$.
Assuming $d(\bar{\alpha})/dt = \lambda \bar{\alpha}$ for some scalar $\lambda$, we have
\begin{equation}
\label{eq: evenlyspacedbar}
(\lambda + \lambda_0)\bar{\alpha} = -\lambda_0 \bar{\alpha}_* - \frac{K}{2}b M \delta S
\end{equation}
and
\begin{equation}
\label{eq: evenlyspacedbarconj}
(\lambda + \lambda_0)\bar{\alpha}_* = -\lambda_0 \bar{\alpha} - \frac{K}{2}b r_0^2 M \delta S,
\end{equation}
where 
\begin{equation}
\label{eq: lambdanought}
\lambda_0 = \frac{K}{2} r_0^2 (1+b).
\end{equation}
We also assume $d(\delta \alpha_M)/dt = \lambda \delta \alpha_M$, so that 
\begin{equation}
\label{eq: alphaM}
(\lambda - \lambda_0)\delta \alpha_M = -\frac{K b}{2} \delta S.
\end{equation}
Because $\delta S$ is the sum of the perturbations, we have
\begin{equation}
\label{eq: evendeltaS}
\delta S = \bar{\alpha} + \delta \alpha_M.
\end{equation}
Substituting this into Eq.\ \eqref{eq: alphaM} and rearranging terms, we get
\begin{equation}
\label{eq: whatever}
\bar{\alpha} + \delta \alpha_M = \bar{\alpha}\left(\frac{\lambda - \lambda_0}{\lambda - \lambda_0 + K b/2}\right).
\end{equation}
Replacing $\delta S$ with \eqref{eq: whatever} in Eqs.\ \eqref{eq: evenlyspacedbar} and \eqref{eq: evenlyspacedbarconj} yields
\begin{equation}
\label{eq: bar2}
\left[(\lambda + \lambda_0) + \frac{K b M}{2} \frac{\lambda - \lambda_0}{\lambda - \lambda_0 + K b/2}\right] \bar{\alpha} = -\lambda_0 \bar{\alpha}_*
\end{equation}
and
\begin{equation}
\label{eq: bar2conj}
(\lambda + \lambda_0) \bar{\alpha}_* = \left(-\lambda_0 + \frac{K b M}{2}r_0^2 \frac{\lambda-\lambda_0}{\lambda - \lambda_0 + K b/2}\right) \bar{\alpha}.
\end{equation}
Combining these two equations and collecting powers of $\lambda$ gives
\begin{equation}
\label{eq: evencubic}
\lambda^3 + \lambda^2 \left(\frac{K b}{2} (M+1) + \lambda_0 \right) + \lambda \left(K b \lambda_0 + \frac{K b}{2} M \alpha_0^2 \lambda_0 - 2 \lambda_0^2 \right) - \frac{K b}{2} M \lambda_0^2 (1 + \alpha_0^2) = 0
\end{equation}

The value of this polynomial at $\lambda = 0$ is $-(K b M/2) \lambda_0^2 (1 + \alpha_0^2)$.  This is a negative number when $K > K_c$, because $\alpha_0^2$ and $\lambda_0$ are positive.  Because the polynomial in Eq.\ \eqref{eq: evencubic} is negative at $\lambda = 0$ and its value tends to positive infinity as $\lambda$ increases, the polynomial must have a root with $\lambda>0$.  The presence of a positive root indicates exponential growth of the perturbations, which means that this equilibrium is unstable.

\end{document}